\documentclass[final,5p,times,twocolumn]{elsarticle}

\newcommand{\gtlt}{\mathrel{\raisebox{0.2ex}{\scriptsize$>$}\!\!\!\raisebox{-0.4ex}{\scriptsize$<$}}}

\usepackage{amssymb}
\usepackage{siunitx}
\usepackage{cuted}

\usepackage{amsfonts,amssymb,amsmath,amsbsy}
\usepackage{physics}
\usepackage{color,calc}
\usepackage{bm}
\usepackage{array}
\usepackage{booktabs}
\usepackage{graphicx}
\usepackage{graphics}
\usepackage{eepic,epsfig}
\usepackage[breaklinks,hidelinks,colorlinks=true,linkcolor=blue,citecolor=blue,filecolor=blue,urlcolor=blue]{hyperref}

\usepackage[dvipsnames]{xcolor}
\definecolor{myblue}{RGB}{0, 0, 255}
\usepackage[toc,page]{appendix}

\usepackage{microtype}

\def\be{ \begin{equation} }
\def\ee{ \end{equation} }
\def\bea{ \begin{eqnarray} }
\def\eea{ \end{eqnarray} }
\def\bse{ \begin{subequations} }
\def\ese{ \end{subequations} }
\def\ba{ \begin{array} }
\def\ea{ \end{array} }
\def\bt{ \begin{tabular} }
\def\et{ \end{tabular} }
\def\beg{ \begin{gathered} }
\def\eeg{ \end{gathered} }

\long\def\/*#1*/{}

\newcommand{\rom}[1]{\expandafter\romannumeral #1\relax}

\journal{Radiation Physics and Chemistry}

\begin{document}

\begin{frontmatter}



\title{FEL Gain Enhancement in an Optical Klystron}

\author[inst1,inst2]{Anahit H. Shamamian}

\affiliation[inst1]{organization={Department of Exact Subjects, Military Academy after Vazgen Sargsyan MoD RA},
            addressline={155 Davit Bek St.}, 
            city={Yerevan},
            postcode={0090}, 
            country={Armenia}}

\affiliation[inst2]{organization={Matinyan Center for Theoretical Physics, A.I. Alikhanyan National Science Laboratory (Yerevan Physics Institute)},
            addressline={2 Alikhanyan Brothers St.}, 
            city={Yerevan},
            postcode={0036}, 
            country={Armenia}}

\author[inst3,inst4]{Hayk L. Gevorgyan}
\cortext[cor1]{Corresponding author}
\ead{hayk.gevorgyan@aanl.am}

\affiliation[inst3]{organization={Experimental Physics Division, A.I. Alikhanyan National Science Laboratory (Yerevan Physics Institute)},
            addressline={2 Alikhanyan Brothers St.}, 
            city={Yerevan},
            postcode={0036}, 
            country={Armenia}}
            
\affiliation[inst4]{organization={Quantum Technologies Division, A.I. Alikhanyan National Science Laboratory (Yerevan Physics Institute)},
            addressline={2 Alikhanyan Brothers St.}, 
            city={Yerevan},
            postcode={0036}, 
            country={Armenia}}

\author[inst2]{Lekdar A. Gevorgian}

\begin{abstract}

A system comprising two identical helical or planar undulators separated by a gap --- viz., an optical klystron (OK) --- is investigated. A formula for the frequency distribution of spontaneous radiation at zero angle is derived. It is shown that the spontaneous radiation line shape gradually narrows with increasing distance (up to an optimal value) between the undulators due to the constructive interference of the radiation fields formed in each of them, while the number of radiated photons decreases. The free-electron laser (FEL) gain coefficient also gradually increases, since it is proportional to the derivative of the spontaneous radiation line shape. Using the undulator parameters of the SASE XFEL and the bunch parameters of the LCLS, the total gain coefficient in the X-ray range is on the order of 2.5 and, for a different electron energy, reaches approximately 90 in the water window frequency range.

\end{abstract}



\begin{keyword}
optical klystron \sep undulator radiation \sep constructive interference \sep gain coefficient \sep stimulated radiation \sep free electron laser
\PACS 0000 \sep 1111
\MSC 0000 \sep 1111
\end{keyword}

\end{frontmatter}


\section{Introduction}\label{Sec:intro}

The peculiarities of FELs distinguish them from other classical electronic devices, making them one of the most promising sources of short-wavelength radiation \cite{Hopf1976, Fedorov1981, Varfolomeev1980, Rukhadze1983, Colson1985}. The core mechanism of an FEL is the generation of stimulated undulator radiation \cite{Madey1971}. The advancement of electron accelerators with sufficiently small angular divergence, low energy spread, high energy, and relatively large current density in the electron beam has significantly contributed to the development of both the theoretical framework and experimental implementation of stimulated radiation from relativistic particles.

In 1977, Vinokurov and Skrinsky proposed a modification of FEL, specifically the OK, which consists of a system of two undulators separated by a gap \cite{vinokurov1977, vinokurov1977a, vinokurov1977b}. An expression for the line shape of spontaneous radiation generated in a single undulator was obtained in \cite{Gevorgyan2025l}, while that for an OK was derived in \cite{GevorgianShamamian1988,ShamamianThesis2016}. In these latter works, it was shown that the line shape in OK narrows due to the interference of the radiation fields emitted by the individual undulators. Consequently, the number of radiated photons decreases proportionally. This problem has also been studied in the works \cite{Dattoli1999,Colson1990,Dattoli2000,Dattoli2000book}. According to Madey's theory \cite{Madey1979}, the gain coefficient of stimulated radiation depends on the derivative of the line shape of spontaneous radiation. Therefore, the narrowing of the line shape leads to an increase in the gain coefficient \cite{GevorgianShamamian1988,ShamamianThesis2016}.

In the studies \cite{Gevorgian2005, HGevorgyan2021b}, X-ray radiation generated by a relativistic charged particle in a crystalline undulator with sections (CUS) was investigated. CUS consists of a system of monocrystals with specified lengths and curvature radii, symmetrically arranged at certain distances from each other. The gain coefficient calculated in \cite{GevorgianShamamian1988,ShamamianThesis2016} corresponds to the values of gain coefficients reported in \cite{Gevorgian2005, HGevorgyan2021b}. For practical applications of the observed effect, it is necessary to determine the optimal values of OK parameter that characterizes the distance between the undulators.

This paper is devoted to the study of the total gain of OK FEL due to constructive interference of radiation fields formed in individual helical or planar undulators. Particularly, the total gains in the case of SASE XFEL undulator and LCLS electron bunch in the X-ray and water window wavelengths of radiation are calculated.

The paper is organized as follows. In Sec.~\ref{Sec:radfield}, an analytical formula is derived for the total radiation field produced by an electron in a system of undulators separated by a gap determined by the line shape function of spontaneous radiation at zero angle. In Sec.~\ref{Sec:Number of Photons}, the number of photons radiated by a single electron oscillating in the helical or planar undulators is calculated. Sec.~\ref{Sec:GainOK} is dedicated to the gain of stimulated radiation formed in the OK. The numerical calculations and results for total gain coefficients in the X-ray and water-window regimes are shown in Sec.~\ref{Sec:GainXWater}. Finally, Sec.~\ref{Sec:Concl} presents the conclusions.

\section{Total Radiation Field and Radiation Line Shape in OK}\label{Sec:radfield}

\begin{figure}[t]
\bt{r}
\centerline{\includegraphics[width=1.0\columnwidth]{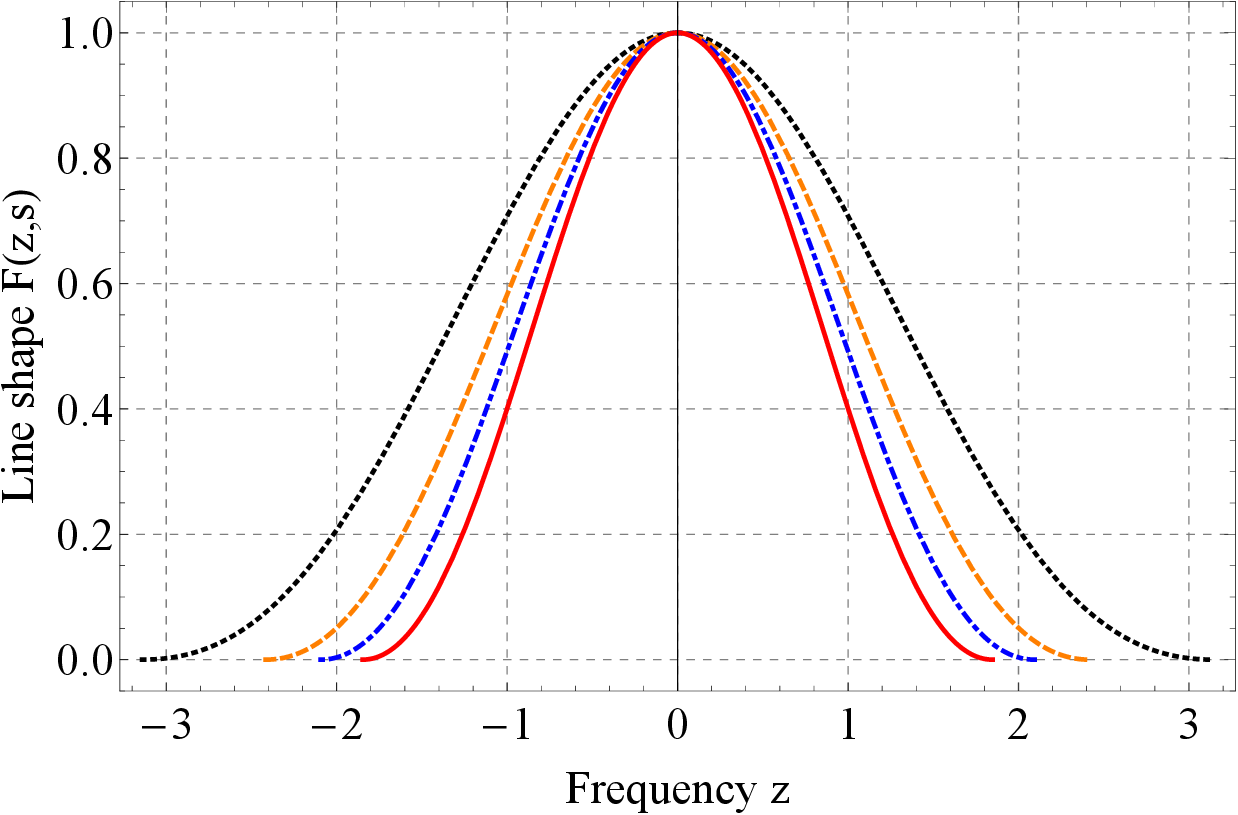}} \\ \\
\centerline{\includegraphics[width=1.0\columnwidth]{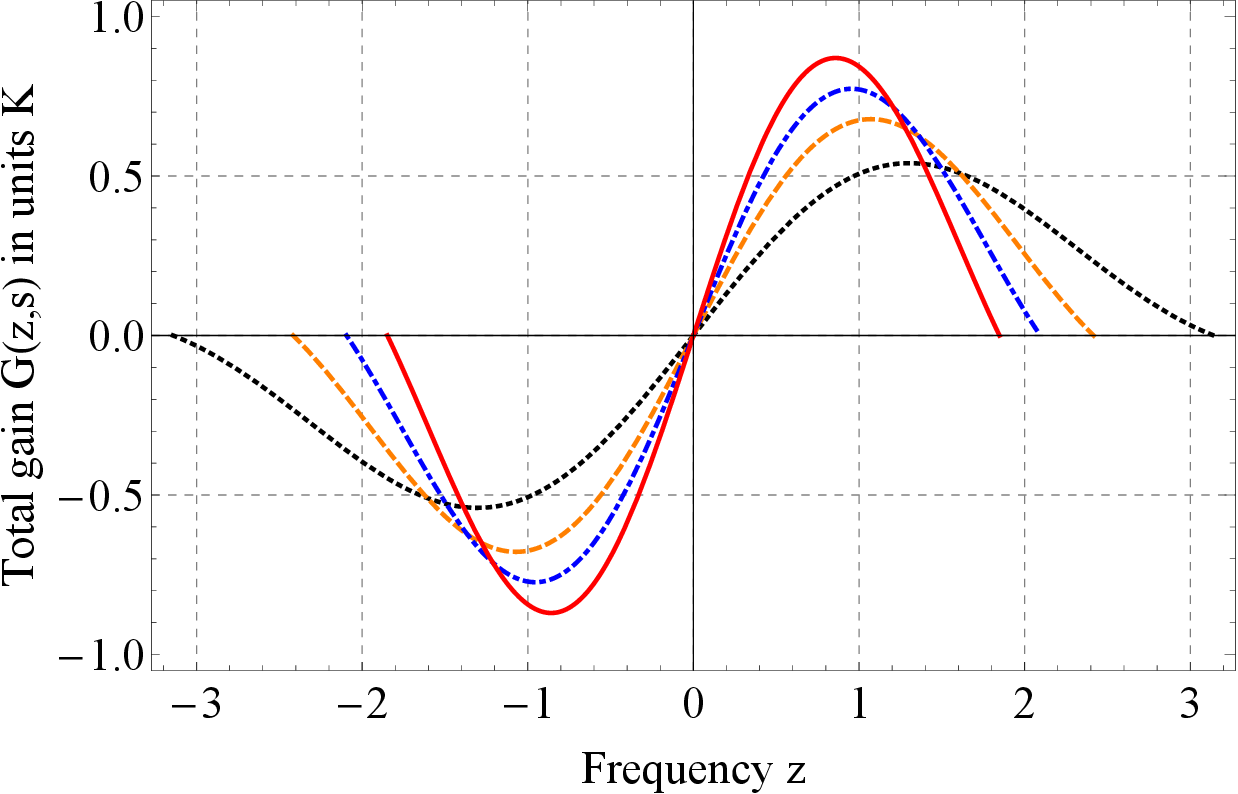}}
\et
\caption{
Line shape $F(z,s)$ (top) and total gain coefficient $G_{h,p} (z,s) \propto - \tfrac{\partial F(z,s)}{\partial z}$ in units $K$ (bottom) as functions of the dimensionless frequency $z$, for different gap values between two undulators: $s=0$ (black dotted), $s=0.3$ (orange dashed), $s=0.5$ (blue dot-dashed), and $s=0.7$ (red). The results apply to both helical (h) and planar (p) undulators (see Eqs.~\eqref{Line shape} and~\eqref{Gain}).
}
\label{fig:lineshape and gain}
\end{figure}

Let's consider the motion of a relativistic electron in a system consisting of two identical helical undulators, each with a length $L$, separated by a gap of $s L$ (where $0<s<1$). Within the undulators, the particle follows a helical trajectory with a longitudinal velocity $\beta_\parallel \, c$ along the $z$-axis of a helical undulator and a transverse velocity $\vb*{\beta}_\perp (t) \, c$ in the $x$-$y$ transverse plane, where $c$ is the speed of light in vacuum. In the gap between the undulators, the particle moves rectilinearly and uniformly.

In the magnetic field of a helical undulator, the particle moves with a velocity in $c$ units given by:
\be
\vb*{\beta} (t) = - \vb*{i} \beta_\perp \sin{\Omega t} + \vb*{j} \beta_\perp \cos{\Omega t} + \vb*{k} \beta_{\parallel},
\ee
where $\vb*{i}$, $\vb*{j}$ and $\vb*{k}$ are the unit vectors along the Cartesian $x$-, $y$-, and $z$-axes, respectively, $\Omega = 2\pi \beta_\parallel c/l$ is the angular frequency of rotation, and $l$ is the spatial period of the undulator.

Neglecting small losses due to radiation primarily caused by transverse motion, the particle’s energy remains constant. Therefore, we assume that the particle moves with a constant average longitudinal velocity, given by $\beta_\parallel = \sqrt{\langle\beta_\parallel (t)^2\rangle} = \sqrt{\beta^2 - \beta_\perp^2}$, where $\beta$ is the initial velocity of the particle. Therefore, $\gamma^2 = \gamma^2_\parallel Q_h$, $Q_h = 1+q^2$, $q = \beta_\perp \gamma$ is an helical undulator parameter (the maximum deflection angle of a charged particle in an undulator), $\gamma$ and $\gamma_\parallel$ are the full and longitudinal (motion) Lorentz factors, respectively. It should be noted that for a planar undulator, $Q_p = 1+q^2/2$.

The trajectory of the electron follows a helical path with a radius $R = \beta_{\perp} c/\Omega$ given by

\be\label{r(t)}
\vb*{r}(t) =
\begin{cases}
\left\{R \cos(\Omega t), R \sin(\Omega t), \beta_{\parallel} c t\right\} \\ \text{for } \, 0 < t < T, \, T (s+1) < t < T (s+2) , \\
\left\{0, 0, \beta_{\parallel} c t\right\} \, \text{for } \, T < t < T (s+1) ,
\end{cases}
\ee
where $T = L/(\beta_\parallel c)$ is its transit time through a single undulator.

The process of photon radiation in undulators is described with sufficient accuracy by classical theory. 
The radiation fields in the first and second undulators are determined using the following integrals 
\bse
\begin{align}
\vb*{I}_1 &= \int\limits_{0}^{T} e^{i(\omega t - \vb*{k} \cdot \vb*{r})} \vb*{a}(t) \, dt ,\\
\vb*{I}_2 &= \int\limits_{(s+1)T}^{(s+2)T} e^{i(\omega t - \vb*{k} \cdot \vb*{r})} \vb*{a}(t) \, dt, \\
\vb*{a}(t) &= \left[\vb*{n} \times\left[\vb*{n} \times \vb*{\beta}(t)\right]\right]. \notag
\end{align}
\ese
Meanwhile, $\omega$ is the frequency and $\vb*{k} = \omega \vb*{n}/c$ is the wave vector of the photon radiated by the electron, $\vb*{n}$ is directional unit vector of radiation given by
\be
\vb*{n} = \{\sin{\vartheta} \cos{\varphi}, \sin{\vartheta} \sin{\varphi}, \cos{\vartheta}\}
\ee
with $\vartheta$ and $\varphi$ being the polar and azimuthal angles of radiation, respectively.

For the frequency-angular distribution of the number of radiated photons in a system of two undulators separated by a gap, we have
\be 
\frac{d^2 N}{d\omega dO} = \frac{\alpha \omega}{4\pi^2} \abs{\vb*{I}}^2, \quad \vb*{I} = \vb*{I}_1 + \vb*{I}_2,  
\ee
where $\alpha = e^2/(\hbar c) = 1/137$ is the fine-structure constant, $d O = \sin{\vartheta} d\vartheta d\varphi$ is the solid angle of radiation at the frequency $\omega$, $e$ is the electron charge, and $\hbar$ is the reduced Planck constant.

Since we are interested in the radiation field formed at the angle $\vartheta= 0$, we have $\vb*{a}(t) = \vb*{\beta}_\perp (t) = 0.5 \beta_\perp \left((1+i) e^{i \Omega t} + (1-i) e^{-i \Omega t} \right)$, and $\omega t - \vb*{k} \cdot \vb*{r} (t) = \omega (1- \beta_\parallel) t$. It is convenient to introduce a dimensionless frequency $x = \omega / \omega_m$, where $\omega_m = 2 \Omega \gamma_\parallel^2 = 2 \Omega \gamma^2 / Q_h$ is the peak frequency of the spontaneous radiation line at zero angle. Let us denote the integration variable by $\omega t = \tau$. Then the integrand takes the form

\be
\frac{\beta_\perp}{2} \left( \left(\vb*{j} - i \vb*{i} \right) e^{i (x-1) \tau} + \left(\vb*{j} + i \vb*{i} \right) e^{i (x+1) \tau} \right)
\ee

The line shape of the spontaneous radiation is formed by the first term, which corresponds to the energy-momentum conservation in the radiation process. Thus, we have 

\be
\beg
\vb*{I}_1 = \vb*{b} J_1, \quad \vb*{I}_2 = \vb*{b} J_2, \quad \vb*{b} = \frac{\beta_\perp}{2\Omega} (\vb*{j} - i \vb*{i}), \\
J_1 = \int\limits_{0}^{2\pi n} e^{i (x-1) \tau} \, d\tau = 2 e^{i \pi n (x-1)} \frac{\sin{\left(\pi n (x-1)\right)}}{x-1}, \\
J_2 = \int\limits_{2\pi n (1+s)}^{2\pi n (2+s)} e^{i (x-1) \tau} \, d\tau = 2 e^{i \pi n (3 + 2 s) (x-1)} \frac{\sin{\left(\pi n (x-1)\right)}}{x-1}, \\
J_1 + J_2 = 2 e^{i \pi n (2+s)} \frac{\sin{\left(2\pi n (x-1)\right)}}{x-1} \cdot \frac{\cos{\left(\pi n (1+s) (x-1)\right)}}{\cos{\left(\pi n (x-1)\right)}}. 
\eeg
\ee
Here, $n = L / l = $ is the number of oscillations performed by the electron in a single undulator. Taking into account that $\abs{\vb*{b}}^2 = \beta_\perp^2/(2\Omega^2)$ for a helical undulator and $\abs{\vb*{b}}^2 = \beta_\perp^2/(4\Omega^2)$ for a planar undulator, accordingly in the case of helical (h) and planar (p) undulators, we obtain 
\be
\lvert \vb*{I}_{h,p} \rvert^2 = \lvert \vb*{I}_1 + \vb*{I}_2 \rvert^2 = \frac{2 \beta^2_\perp}{\Omega^2} F(x,s) \cdot \left\{1, \frac{1}{2}\right\}, 
\ee
where $F(x,s)$ is the line shape function of spontaneous radiation at zero angle,
\be
F(x,s) = \frac{\sin^2{(2\pi n (x-1))}}{(x-1)^2} \cdot \frac{\cos^2{(\pi n (1+s) (x-1))}}{\cos^2{\left(\pi n (x-1)\right)}}.
\ee

The quantity $F(x,s)$ is a sharply peaked function of $x$, meaning that the probability density of spontaneous radiation at zero angle decreases rapidly as $x$ deviates from the peak frequency $x=1$.

\section{Probability of Spontaneous Radiation and the Number of Radiated Photons}\label{Sec:Number of Photons}
The frequency distribution of photons radiated at zero angle is given by the following expression:
\be
\beg
\frac{dN_{h,p}}{d\omega} = \frac{\alpha \omega}{4\pi^2} \abs{\vb*{I}_{h,p}}^2 \int \, dO , 
\eeg
\ee
where the total solid angle element at zero radiation angle (in the upper or forward hemisphere) is $\int \, dO = \int_{0}^{2\pi} \int_{0}^{\pi/2} \sin{\theta} \, d\theta \, d\varphi = 2\pi$, where the radiation angle is $\theta = \vartheta \gamma$ for undulator radiation from relativistic particles.

The radiation probability with respect to the variable $x$ has the form:
\be
\begin{gathered}
\frac{dN_{h,p}}{dx} = \frac{dN_{h,p}}{d\omega} \frac{d\omega}{dx} = \frac{4 \alpha q^2}{\pi Q_{h,p}} x F(x,s) \cdot \left\{1, \frac{1}{2}\right\}, \\
1 - \Delta \leq x \leq 1 + \Delta,  \quad \Delta \overset{\Delta}{=} \frac{1}{2 n (1+s)}, \quad n \gg 1 . \\
\end{gathered}
\ee

Since $F(1,s) = 4 \pi^2 n^2$ and $F(1 \pm \Delta) = 0$, i.e. the width of the Dirac $\delta$-like function $F(x,s)$ is $\Delta$, then, to within terms of order $\Delta$, we have
\be
F(x,s) = \frac{2\pi^2 n}{1+s} \delta(x-1).
\ee

The number of photons radiated by a single electron oscillating in the undulator is
\be
N_{h,p} = \frac{8\pi \alpha n q^2}{(1+s) Q_{h,p}} \cdot \left\{1, \frac{1}{2}\right\}.
\ee

\section{Gain of Stimulated Radiation Formed in the Optical Klystron}\label{Sec:GainOK}

It is convenient to introduce a new dimensionless variable $z = 2\pi n (x-1)$
\be\label{Line shape}
\begin{gathered}
\frac{dN_{h,p}}{dz} = \frac{8 \alpha n q^2}{Q_{h,p}} \left(1 + \frac{z}{2\pi n}\right) F(z,s) \cdot \left\{1, \frac{1}{2}\right\}, \\
F(z,s) = \frac{\sin^2 z}{z^2} \cdot \frac{\cos^2{\left(z (1+s)/2\right)}}{\cos^2{z/2}},\\
 - \frac{\pi}{1+s} \leq z \leq \frac{\pi}{1+s}.
\end{gathered}
\ee
It should be noted that the spectral width of the line shape $F(z,s)$ is narrowed by a factor of $1+s$ compared to the spectral width of the line $F(z,0)$. Consequently, the derivative of the line shape increases by a factor of $1+s$. Fig.~\ref{fig:lineshape and gain} shows the dependence of $F(z,s)$ on $z$ and the corresponding $- \tfrac{\partial F(z,0)}{\partial z}$ for various values of the gap parameter $s$.

To a good approximation, neglecting terms of order $z/(2\pi n)$, we have:
\be
S_{h,p} (z,s) \overset{\Delta}{=} \frac{dN_{h,p}}{dz} = \frac{8 \alpha n q^2}{Q_{h,p}} F(z,s) \cdot \left\{1, \frac{1}{2}\right\}.
\ee

The cross-section for stimulated radiation or absorption is given by: 
\be
\sigma (z, s) = 2 \pi \lambda^2 S_{h,p} (z,s), 
\ee
where $\lambda$ is the wavelength of the spontaneous radiation. 

The effect of quantum recoil during radiation or absorption leads to a violation of detailed balance. In this case, in the region $z > 0$, the stimulated radiation cross-section becomes larger than the absorption cross-section, while in the region $z < 0$ the situation is reversed. If the quantum recoil effect during photon radiation or absorption by an electron is neglected, the cross-sections for these processes coincide. When quantum recoil is taken into account, a frequency shift in $x$ or $z$ occurs:

\be\label{deltaz}
\frac{\Delta z}{z} = \mp \frac{\hbar \omega}{m c^2 \cdot \gamma} = \mp \frac{2 \lambda_c}{\lambda \gamma}, \quad \left( z \gtlt 0\right).
\ee
where $\lambda_c = h / (m c) \approx 2.426 \times 10^{-10}$\SI{}{\centi\meter} is the Compton wavelength of the electron.

In the process of stimulated radiation or absorption, photons with different wavelengths participate. Stimulated radiation occurs at wavelengths $\lambda < \lambda_0$, where $\lambda_m$ is the peak wavelength of the line shape at which no stimulated process occurs. The probability of spontaneous radiation in the region $\lambda < \lambda_m$ is equal to $0.5$. Consequently, the gain coefficient for stimulated radiation per unit path length of an electron bunch with density $\rho$ \cite{Marshall1985} is equal:

\be
g_{h,p}(z,s) = \pi \lambda^2 \rho \, \delta S_{h,p} (z,s) = \frac{8 \pi \alpha \lambda^2 \rho n q^2}{Q_{h,p}} \frac{\partial F(z,s)}{\partial z} \, \Delta z \cdot \left\{1, \frac{1}{2}\right\}, 
\ee
where, taking into account Eq.~\eqref{deltaz}, $\tfrac{\partial F(z,s)}{\partial z} \, \Delta z = - \tfrac{2 \lambda_c}{\lambda \gamma} \tfrac{\partial F(z,s)}{\partial z}$, the total (over the full undulator length) gain is:

\be\label{Gain}
\begin{gathered}
G_{h,p} (z,s) = 2 n l \cdot g_{h, p} (z, s) = K \left(- \frac{\partial F(z,s)}{\partial z} \right), \\
K = \frac{16 \pi \alpha \lambda_c n^2 l^2 q^2}{\gamma^3} \cdot \left\{1, \frac{1}{2}\right\}.
\end{gathered}
\ee
The quantity $- \tfrac{\partial F(z,0)}{\partial z}$ reaches its maximum value of $0.54$ at $z_m = 1.303$ (see Fig.~\ref{fig:lineshape and gain}) \cite{Marshall1985}, and the value of $- \tfrac{\partial F(z,s)}{\partial z}$ is increased by a factor of $1+s$. Taking into account $\abs{\Delta z_m} = z_m \left(\tfrac{2 \lambda_c}{\lambda \gamma}\right)$, the maximum value of $g_{h,p} (z,s)$ is:
\be
\max_z g_{h,p} (z, s) \overset{\Delta}{=} g_{h,p} (z_m, s) = 6.262939 \times 10^{-11} \left\{1, \frac{1}{2}\right\} \cdot \rho \frac{n q^2 \lambda (1+s)}{Q_{h,p} \gamma}.   
\ee
The maximum value of the total gain coefficient is $\max G_{h,p} = 2 n l \cdot \max_z g_{h,p} (z, s) = 1.2525878 \times 10^{-10} \cdot \tfrac{\rho n^2 q^2 l \lambda}{Q_{h,p} \gamma} (1+s)$.



\section{Gain in the X‑Ray and Water‑Window Regimes}\label{Sec:GainXWater}
Let us consider OK with parameters of an SASE XFEL undulator and an electron bunch from the LCLS. The undulator parameters are as follows: spatial period $l = 3$\SI{}{\centi\meter}, number of oscillations in a single undulator $n=113$, field parameter $\beta_\perp = 1.3105895 \times 10^{-4}$, $Q_p = 1 + q^2/2$, where $q = \gamma \beta_\perp$.
The parameters of the electron bunch are: total number of electrons $N_b =1.56 \times 10^9$, standard deviations in the transverse and longitudinal directions $\sigma_R = 6.12 \times 10^{-4}$\SI{}{\centi\meter} and $\sigma_L = 9\times 10^{-4}$\SI{}{\centi\meter}, respectively. The average electron density in the bunch can be estimated as $\rho = N_b/(2\pi \sigma_R^2 \sigma_L) = 7.365\times 10^{17}$.

\begin{enumerate}
\item \textbf{X-ray range:} \\

For $\max G_{h,p}$ at a wavelength of $\lambda = 1.5$\SI{}{\angstrom} ($\gamma = 2.6614476 \times 10^4$) we obtain: 
\be
\max G_{h,p} = 1.6714619 (1+s_0) = 2.4403343,
\ee
where $s_0 = 0.46$ is the optimal value of $s$, which is determined by using a special method: it cannot be obtained by strict mathematical optimization, since the necessary extremum condition for a function of two variables is not satisfied.

\item \textbf{Water window range:} \\
At a wavelength of $\lambda_w = 4.2 \times 10^{-7}$\SI{}{\centi\meter} a bunch with energy $\gamma_w = 1.919486 \times 10^3$ must be used. This gives $q^2_w = \gamma^2_w \beta^2_\perp = (2.525842 \times 10^{-1})^2 = 6.3798778 \times 10^{-2}$, $Q_p (\gamma_w) = 1.03189939$. For the same optimal value  $s_0 = 0.46$, we obtain:
\be
\max G_{h,p} = 89.126716.
\ee
\end{enumerate}
The values of $\gamma$ and $\gamma_w$ are determined from the formula $\gamma = \gamma_\parallel/ \sqrt{1- q_\parallel^2/2}$, where $q_\parallel = \gamma_\parallel \beta_\perp$, and $\gamma_\parallel = \sqrt{l/(2\lambda_{,w})}$. 

\section{Conclusion}\label{Sec:Concl}
We have considered the process of stimulated radiation in an OK. Expressions for the gain coefficients have been obtained for both helical and planar undulators. The total gain coefficients of stimulated radiation generated in a planar undulator with known parameters have been calculated for both the X-ray and water window frequency ranges. In the X-ray range, it is on the order of 2.5, and in the water window it reaches approximately 90. Moreover, the photon beams are more monochromatic in both cases.

It is shown that the gaps between undulators in SASE XFEL setups --- used for restoring the electron bunch parameters --- lead, due to constructive interference of the radiation fields emitted by individual undulators, to a narrowing of the spontaneous radiation line shape, with a corresponding reduction in the number of radiated photons. However, as a result of the gain of stimulated radiation in the FEL process, a photon beam with improved monochromaticity is produced.

 \bibliographystyle{elsarticle-num} 


\end{document}